\documentclass[superscriptaddress,twocolumn,preprintnumbers,amsmath,amssymb]{revtex4-2}

\usepackage{graphicx}
\usepackage{dcolumn}
\usepackage{bm}
\usepackage{natbib}
\usepackage{soul}

\makeatletter
\def\@fnsymbol#1{%
   \ifcase#1\or
   \TextOrMath \textdagger \dagger\or
   \TextOrMath\textasteriskcentered *\or
   \TextOrMath \textdaggerdbl \ddagger \or
   \TextOrMath \textsection  \mathsection\or
   \TextOrMath \textparagraph \mathparagraph\or
   \TextOrMath \textbardbl \|\or
   \TextOrMath {\textdagger\textdagger}{\dagger\dagger}\or
   \TextOrMath {\textdaggerdbl\textdaggerdbl}{\ddagger\ddagger}\else
   \@ctrerr \fi
}
\makeatother

\begin{document}

\title{Anyon braiding and telegraph noise in a graphene interferometer}

\author{Thomas Werkmeister}
\thanks{These authors contributed equally to this work}
\affiliation{
John A. Paulson School of Engineering and Applied Sciences, Harvard University, Cambridge, MA 02138, USA
}
\author{James R. Ehrets}
\thanks{These authors contributed equally to this work}
\affiliation{
Department of Physics, Harvard University, Cambridge, MA 02138, USA
}
\author{Marie E. Wesson}
\affiliation{
John A. Paulson School of Engineering and Applied Sciences, Harvard University, Cambridge, MA 02138, USA
}
\author{Danial H. Najafabadi}
\affiliation{
Center for Nanoscale Systems, Harvard University, Cambridge, MA 02138, USA
}
\author{Kenji Watanabe}
\affiliation{
Research Center for Electronic and Optical Materials, National Institute for Materials Science, 1-1 Namiki, Tsukuba 305-0044, Japan
}
\author{Takashi Taniguchi}
\affiliation{
Research Center for Materials Nanoarchitectonics, National Institute for Materials Science,  1-1 Namiki, Tsukuba 305-0044, Japan
}
\author{Bertrand I. Halperin}
\affiliation{
Department of Physics, Harvard University, Cambridge, MA 02138, USA
}
\author{Amir Yacoby}
\affiliation{
John A. Paulson School of Engineering and Applied Sciences, Harvard University, Cambridge, MA 02138, USA
}
\affiliation{
Department of Physics, Harvard University, Cambridge, MA 02138, USA
}
\author{Philip Kim}
\thanks{Corresponding author. Email: pkim@physics.harvard.edu}
\affiliation{
John A. Paulson School of Engineering and Applied Sciences, Harvard University, Cambridge, MA 02138, USA
}
\affiliation{
Department of Physics, Harvard University, Cambridge, MA 02138, USA
}

\date{\today}
\begin{abstract}
The search for anyons, quasiparticles with fractional charge and exotic exchange statistics, has inspired decades of condensed matter research. Quantum Hall interferometers enable direct observation of the anyon braiding phase via discrete interference phase jumps when the number of encircled localized quasiparticles changes. Here, we observe this braiding phase in both the $\nu = 1/3$ and $4/3$ fractional quantum Hall states by probing three-state random telegraph noise (RTN) in real-time. We find that the observed RTN stems from anyon quasiparticle number $n$ fluctuations and reconstruct three Aharonov-Bohm oscillation signals phase shifted by $2\pi/3$, corresponding to the three possible interference branches from braiding around $n$ (mod 3) anyons. Hence, we reveal the anyon braiding phase using new methods that can readily extend to non-abelian states.
\end{abstract}
\maketitle

\subsection*{Introduction}
The fractional quantum Hall (FQH) effects, where electrons are confined to two spatial dimensions and exposed to large magnetic fields, have long been predicted to host emergent fractionally charged excitations that obey neither fermionic nor bosonic exchange statistics \cite{laughlin1983, halperin1984, arovas1984, stern2008, feldman2021}. These quasiparticles fall into two classes, abelian and non-abelian anyons, associated with FQH states at different filling factor $\nu$. For abelian anyons, exchanging two quasiparticles (i.e. braiding) results in the many-body wavefunction acquiring a complex phase, while in the non-abelian case, the wavefunction evolves by a unitary transformation, enabling fault-tolerant topological quantum computation that has yet to be experimentally realized \cite{kitaev2006, nayak2008, stern2013, sarma2015, yazdani2023}.

FQH Fabry-Pérot (FP) interferometers enable direct measurements of the anyon braiding phase. \cite{dec.chamon1997, halperin2011}. By partially backscattering current at two quantum point contact (QPC) constrictions, the conductance through the FP cavity modulates with the phase accumulated by edge-traveling quantum Hall (QH) quasiparticles \cite{feldman2021, carrega2021}. When the number of localized cavity quasiparticles encircled by the interfering edge changes, the resulting phase shift in the interference signal is equivalent to twice the fundamental exchange phase (modulo $2\pi$) \cite{read2024, nakamura2020}. However, since the observed discrete phase jumps represent charging events in the cavity, resultant Coulomb effects on the interference path can pose complications in probing the braiding effect. In particular, the Coulomb coupling between the bulk and edge must be accounted for when extracting the anyon braiding phase from a small number of phase jump events \cite{nakamura2022}. Consequently, mitigating Coulomb coupling while maintaining electrostatic tunability has been a major challenge in semiconductor-based interferometers \cite{nakamura2019, nakamura2020}.
\begin{figure*}
    \centering
    \includegraphics[width = \textwidth]{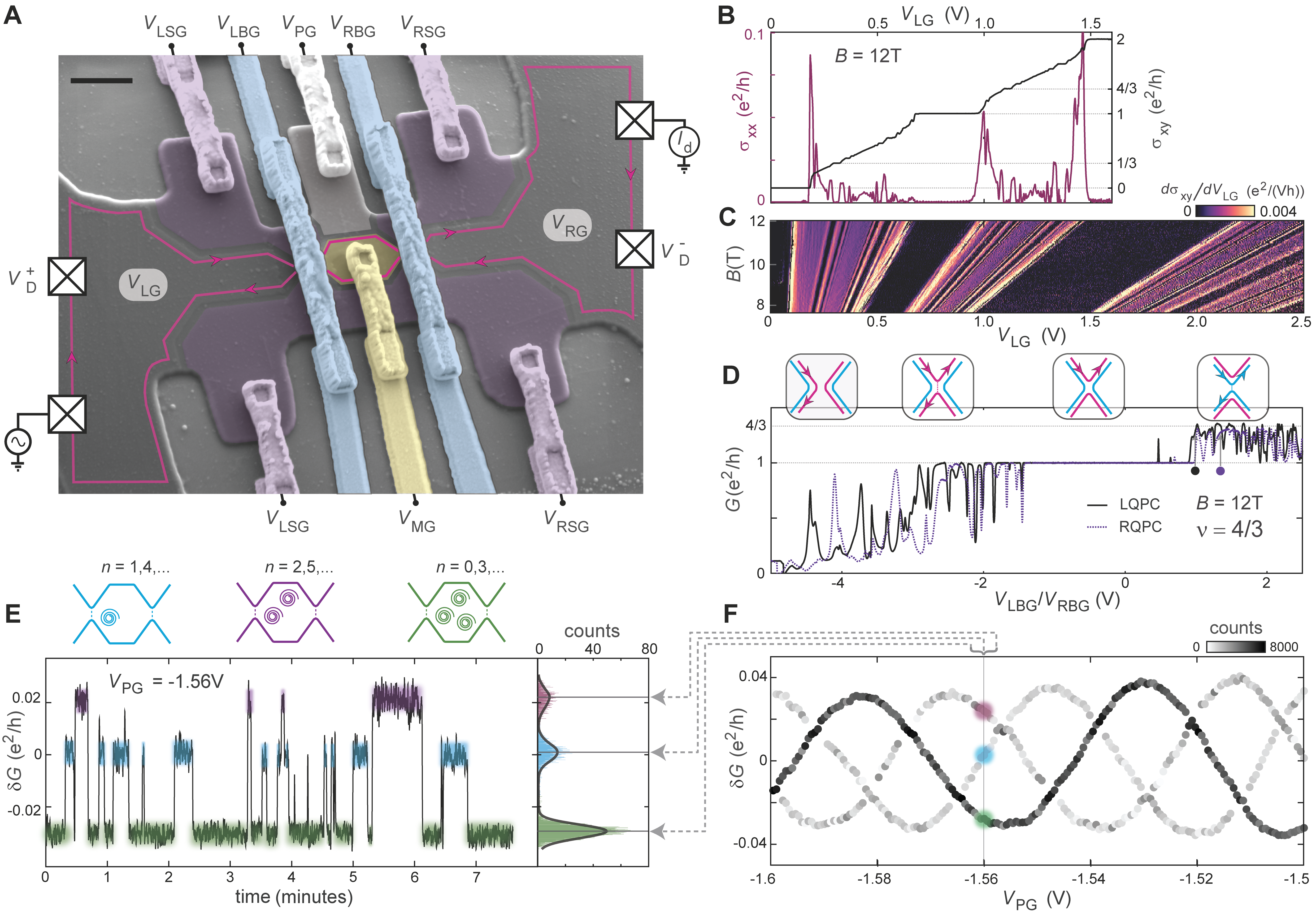}
    \caption{\textbf{Graphene FQH interferometer telegraph noise.} (\textbf{A}) False-color scanning electron microscope image of a representative device depicting an interference measurement with a single edge channel. Scale bar: $1 \mu$m. We measure the conductance $G=I_d/(V_D^+ - V_D^-)$ where $I_d$ is the measured drain current and $V_D^\pm$ are the measured voltages at the indicated locations. The top graphite gate is etched into 8 separately tunable regions controlled by voltages $V_{LG}$, $V_{LSG}$, $V_{PG}$, $V_{MG}$, $V_{RSG}$, and $V_{RG}$. Additionally, voltages $V_{LBG/RBG}$ applied to suspended bridges above the two QPCs independently set their transmissions. See Fig. S1 for more details. (\textbf{B}) Hall conductivity $\sigma_{xy}$ and longitudinal conductivity $\sigma_{xx}$ measured with contacts in the left reservoir of the device highlighting the robust plateaus at $\nu = 0$, $1/3$, $1$, and $4/3$, which are the quantum Hall states we use in this work. (\textbf{C}) Derivative of $\sigma_{xy}$ showing the expected magnetic field dependence of the well-developed quantum Hall states. (\textbf{D}) Conductance $G$ across the left/right QPC tuned by $V_{LBG/RBG}$ with the right/left QPC deactivated and the bulk in $\nu = 4/3$. See Fig. S2 for more details on QPC tunings. Inset: schematics of the edge channel configuration and tunneling at a QPC. As $V_{LBG/RBG}$ increases, the QPC opens more and transmits the outer integer edge (pink) first and eventually transmits the inner fractional edge (blue) with weak backscattering. (\textbf{E}) Three-state RTN in the conductance $G$ when the fractional edge of $\nu = 4/3$ is weakly backscattering (corresponding to the tunings marked by the black/purple dots for the left/right QPC in D). Top inset: cartoons of the three possible interference branches from braiding around $n$ (mod 3) anyons, corresponding to the three conductance levels. Right inset: histogram of the conductance over time with 1000 bins. (\textbf{F}) Extracted sinusoidal oscillations from histograms over approximately 8~minutes for each $V_{PG}$ voltage. Each data point shows the central conductance value of a gaussian fit to a histogram peak, shaded by the total counts summed under the gaussian. See Fig. S3 for similar data in $\nu = 1/3$. All data is at $T = 20$~mK and $B = 12$~T unless otherwise noted.}
    \label{fig1}
\end{figure*}

Graphene-based interferometers offer several advantages for realizing anyon braiding, both for the abelian and non-abelian cases. First, both QPCs \cite{zimmermann2017, overweg2018} and interferometers \cite{deprez2021, ronen2021, zhao2022, fu2023, yang2023} exhibit robust electrostatic tunability for integer QH states. Second, graphite gates reduce Coulomb coupling to reveal Aharonov-Bohm (AB) oscillations \cite{ronen2021}, which previously required complex quantum well structures in GaAs \cite{nakamura2019, nakamura2020}. Simultaneously, these atomically-flat gates enhance FQH states by screening out charge disorder \cite{zeng2019, ribeiro-palau2019} while allowing precise control of QPC transmission \cite{ronen2021, cohen2023, cohen2023a} and FP cavity filling factor \cite{werkmeister2024}. Encouragingly, bilayer graphene has displayed a multitude of even-denominator FQH states which may host non-abelian anyons with relatively large energy gaps \cite{zibrov2017, li2017, huang2022, assouline2024, hu2023}, as required for topologically-protected qubits. Less explored even-denominator states have also been observed in monolayer graphene \cite{zibrov2018, kim2019}.

In this work, we report direct observation of abelian anyon braiding using high-visibility Aharonov-Bohm interference in single atomic layer graphene. Owing to the comprehensive electrostatic control of the device, we are able to tune to both $\nu = 1/3$ and $\nu = 4/3$ at a fixed magnetic field. Strikingly, we observe three-level random telegraph noise (RTN) in each of these FQH states, which was predicted \cite{kane2003, grosfeld2006, rosenow2012} but never observed in previous experiments \cite{nakamura2020, nakamura2022, nakamura2019, nakamura2023, willett2023, kundu2023}. Counterintuitively, we show how fluctuations in anyon number within the cavity enable us to directly observe the anyonic braiding phase with minimal complications due to Coulomb effects.

\subsection*{Constructing a graphene fractional quantum Hall interferometer}
We construct our device by performing a series of previously described nanolithography steps \cite{ronen2021, werkmeister2024} on a monolayer graphene-based van der Waals heterostructure encapsulated with insulating hexagonal boron nitride \cite{dean2010} and conducting graphite gates. By gating with separated top graphite regions, we define the interferometer cavity and construct two QPCs that introduce backscattering between the opposite chirality QH edge channels, as shown in Fig. 1A. The quality of the graphene channel is maintained in part due to the encapsulating graphite gates, as evident by well-developed FQH states (Fig. 1B,C) at $T = 20$~mK. Unlike previous FQH interferometers in GaAs, we can tune the electron density in-situ in graphene to access many FQH states while holding the magnetic field fixed. In this work, we operate at $B = 12$~T (unless noted otherwise) and change density to tune between $\nu = 4/3$, $\nu = 1$, and $\nu = 1/3$, enabling a direct comparison of interference between each state.

We begin by tuning the device such that the interferometer cavity and adjoining contact regions are set to $\nu = 4/3$ by applying appropriate voltages $V_{MG}$, $V_{LG}$, and $V_{RG}$, while we define the cavity boundary by setting the plunger gate ($V_{PG}$)  and QPC split-gate regions ($V_{LSG/RSG}$) to lie within $\nu = 0$. We tune two additional metallic bridge gates over the QPCs (applying $V_{LBG/RBG}$) to independently set the QPC transmissions while keeping all other gates fixed. The conductance across each QPC transitions from near 0 to a plateau at $e^2/h$, and finally to near $(4/3) e^2/h$ with increasing $V_{LBG/RBG}$ as shown in Fig. 1D. The resonances in the conductance in these scans indicate a relatively soft confining potential at this QPC tuning \cite{cohen2023, baer2014}. Other tunings at $\nu = 1/3$ and $\nu = 1$ have fewer resonances, though we find that the general behavior of the interference signal is unaffected by these transmission resonances. It is then straightforward to tune each QPC to weakly backscatter the fractional edge channel of $\nu = 4/3$ by setting $V_{LBG/RBG}$ as indicated in Fig. 1D.

\subsection*{Three-state anyon telegraph noise}
By measuring the conductance through the interferometer with the fractional edge of $\nu = 4/3$ weakly backscattering at both QPCs, we discover sporadic switches between three discrete levels as a function of time i.e. three-state RTN (Fig. 1E). From a histogram showing how these levels are weighted over a 50 minute scan, we extract the average conductance and the total weight in time associated with each level. Then, we step the plunger gate $V_{PG}$ to modify the area \cite{ronen2021, nakamura2019, ofek2010, mcclure2012}, or equivalently the total charge \cite{werkmeister2024, feldman2022} contained in the FP cavity, and repeat the process. Plotting the resulting average conductance values shaded by their weight in time as a function of $V_{PG}$ reveals three interwoven sinusoidal oscillations separated in phase by $2\pi/3$ (Fig. 1F).

Each one of these sinusoids, separated in phase by $2\pi/3$, represent one of the three $n$ (mod 3) possible phase “branches” associated with braiding the interfering edge quasiparticles around a system composed of $n$ $e/3$ abelian anyons \cite{kane2003, grosfeld2006, rosenow2012}.

Demonstrating exchange statistics with RTN is a departure from the typical method used in previous interferometer studies, in which discrete phase jumps in the continuously modulating AB phase $\theta_{AB} (B, V_{PG})$ were argued to represent the addition or subtraction of a quasiparticle in the interferometer bulk \cite{nakamura2020, nakamura2022, nakamura2023}. These studies rely on modulating the AB phase using magnetic field $B$ and plunger gate $V_{PG}$ to probe irregularly spaced phase jumps due to anyon localization. In order to extract the anyon braiding phase, however, the experimentally observed phase jumps need to be analyzed with consideration given to the electrostatic coupling between the QH edge and localized bulk states. These analyses often become complicated as the bulk-edge capacitive coupling, the position of the chemical potential relative to the bulk Landau levels, and interferometer area $A$ can drastically impact interference behavior, such as pivoting the interference signal into the well-studied “Coulomb-dominated” regime \cite{halperin2011, nakamura2022, ofek2010, zhang2009, vonkeyserlingk2015}. It has also been shown that integer quantum Hall fillings can produce similar fractional phase jumps due to the Coulomb coupling effect described above \cite{yang2023, werkmeister2024, roosli2020, roosli2021}.

The measurement scheme presented in our work, i.e. observing a time-dependent interference phase $\phi$ due to the RTN, offers a key benefit in this regard. Here the parameters $(B,V_{PG})$ used to modulate the AB phase 
\begin{equation*}
    \theta_{AB} (B, V_{PG}) = 2\pi \frac{e^*}{e}\frac{A(V_{PG}) B}{\phi_0}
\end{equation*}
can be held fixed while the quasiparticle number $n(t)$ fluctuates in time. Then, the expression for the total interference phase may be simplified as
\begin{equation}
        \phi = \theta_{AB} (B, V_{PG}) + n(B, V_{PG}, t) \ \theta_a \rightarrow \theta_{AB}  + n(t) \ \theta_a
\label{eqn1}
\end{equation}
where $\theta_a = 2\pi/3$ is the braiding phase for an $e^* = e/3$ anyon around a localized counterpart (assuming that Coulomb interactions are well screened) and $\phi_0 = h/e$. Considering $2\pi$ periodicity of $\phi$, the resulting braiding phase can only be one of three values varying by $2\pi/3$, reflecting the total number of localized anyons $n(t)$ (mod 3) at a given time. The device accordingly remains static in configurable parameter space $(B,V_{PG})$ while observing conductance switching from real-time fluctuations $n(t)$. Therefore, as this method demonstrates the existence of all three branches associated with the exchange statistics of charge $e/3$ anyons, we construct a complete representation of the state’s braiding outcomes near a fixed device configuration set by $B$, $V_{PG}$, and density (tuned via $V_{MG}$). Additionally, by sampling a large number of phase jumps while clearly resolving three branches at each $V_{PG}$ tuning, we demonstrate that Coulomb interactions are indeed negligible in our devices, allowing observation of $\theta_{AB}$ from the evolution of the individual branches shown in Fig. 1F at fixed $B$.

\subsection*{Aharonov-Bohm magnetic field dependence}
\begin{figure*}
    \centering
    \includegraphics[width = \textwidth]{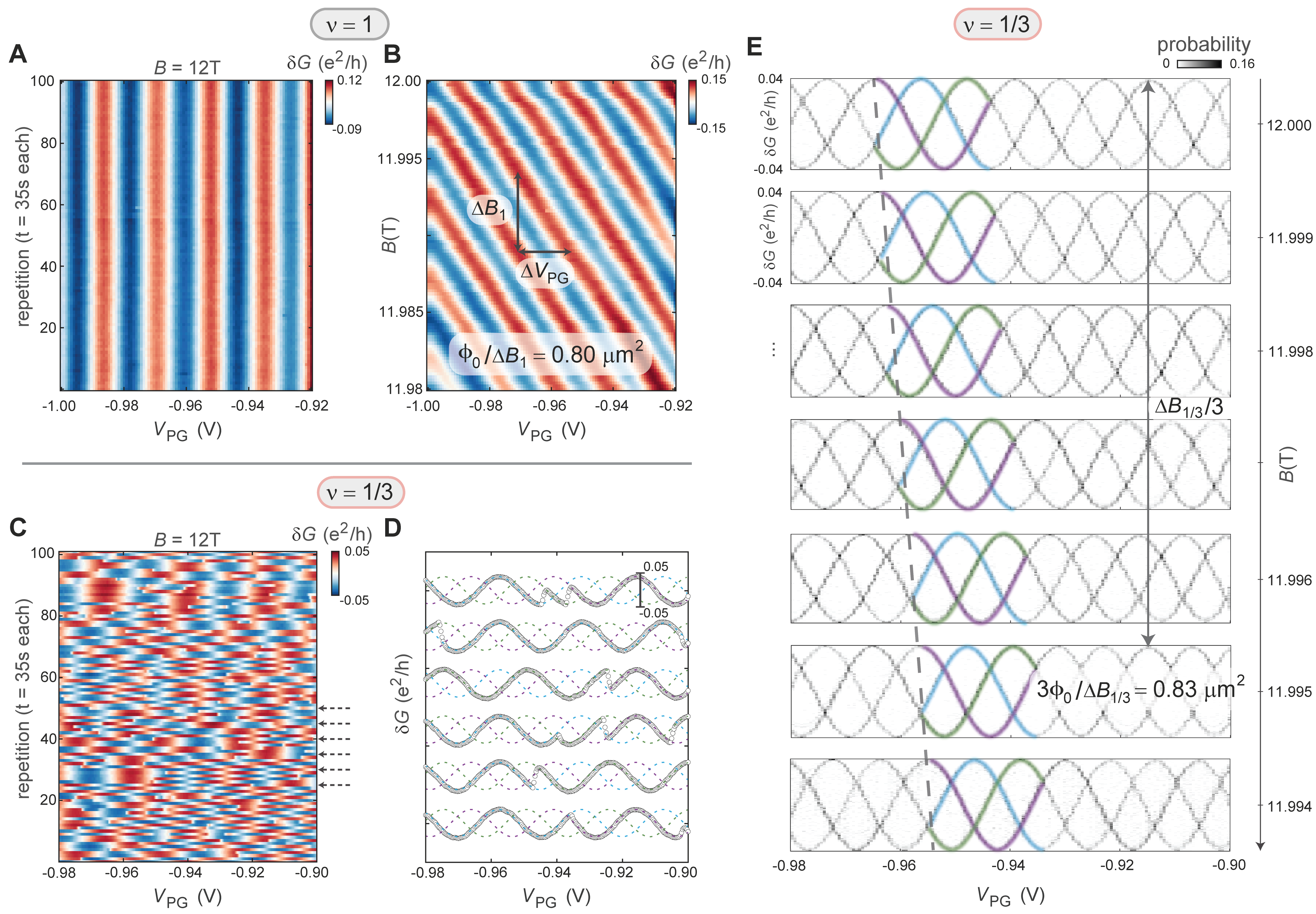}
    \caption{\textbf{Aharonov-Bohm magnetic field trend of $\bm{\nu = 1/3}$ branches.} (\textbf{A}) 100 repeated $V_{PG}$ sweeps in $\nu = 1$, where each sweep takes 35~s, demonstrating the temporal stability of oscillations in the integer state. (\textbf{B}) Magnetic field dependence of the oscillations in $\nu = 1$ demonstrating Aharonov-Bohm periodicity since the magnetic field period yields $\phi_0/\Delta B_1 = 0.80 \ \mu \text{m}^2$, consistent with lateral depletion tending to shrink the area enclosed by the edge state bound by a lithographic design of $1.1 \ \mu \text{m}^2$. (\textbf{C}) 100 repeated $V_{PG}$ sweeps in $\nu = 1/3$, the same sweeping parameters as in (A), demonstrating the stochastic fluctuations induced by the RTN that could easily be mistaken for structureless noise, especially if averaged over by a slow sweep. (\textbf{D}) 6 of the $V_{PG}$ sweeps plotted to demonstrate switching between the three sinusoidal branches within a single scan. (\textbf{E}) 2D histograms showing the probability of measuring a given conductance for each $V_{PG}$ value and its trend with the magnetic field. All three branches (partially highlighted) are made evident by this plot, and each trends identically with a flux super-period in the magnetic field corresponding to $e/3$ anyons.}
    \label{fig2}
\end{figure*}
\begin{figure*}
    \centering
    \includegraphics[width = \textwidth]{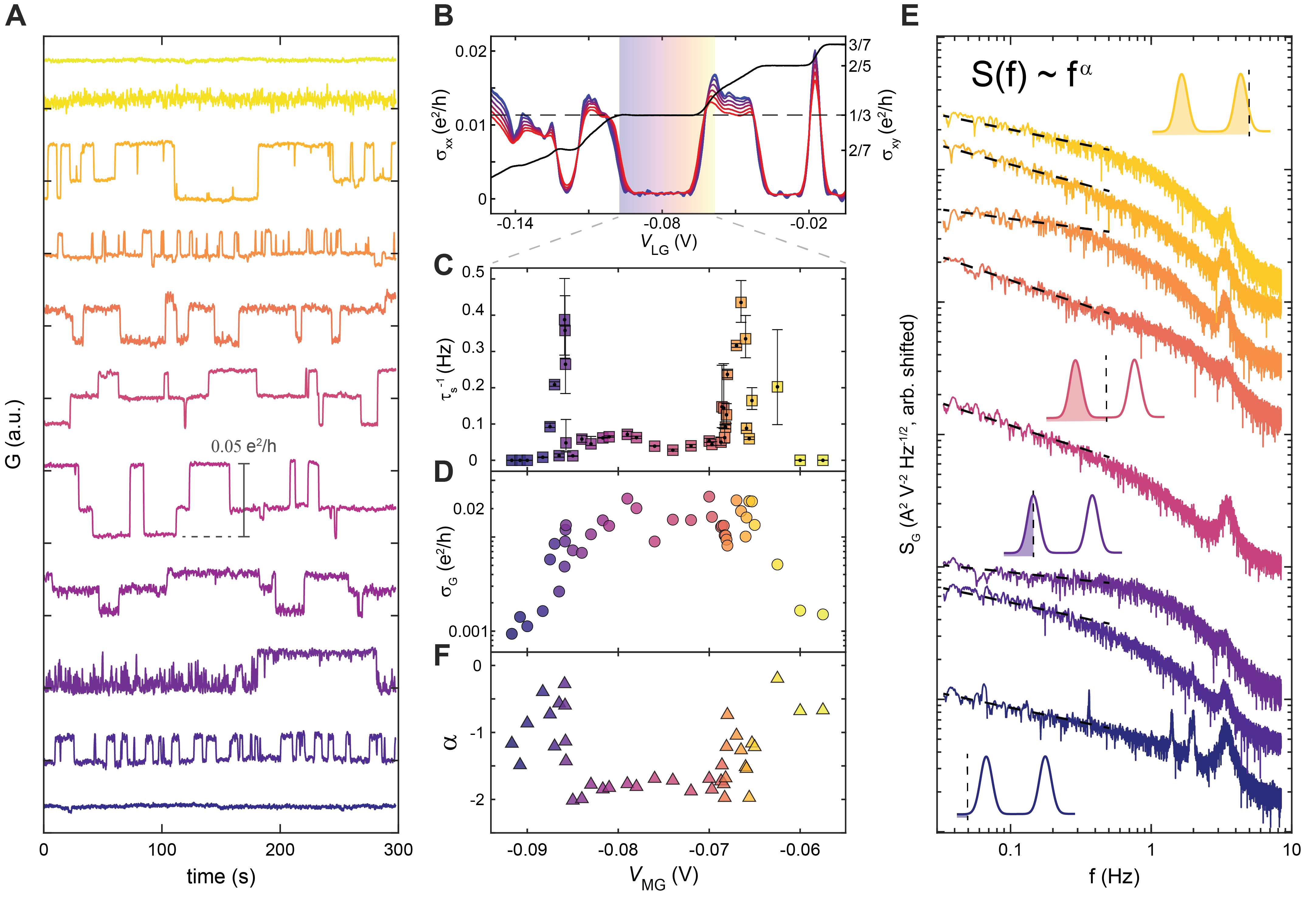}
    \caption{\textbf{Telegraph noise dependence on filling in $\bm{\nu =1/3}$.} (\textbf{A}) Conductance versus time taken at different values for $V_{MG}$,  indicated by color (traces are arbitrarily vertically shifted by ordering of $V_{MG}$). The 300~seconds shown are from longer 50~minute datasets. (\textbf{B}) Hall data taken from the left half of the device with density tuned via $V_{LG}$. The color shade bar indicates the associated range shown for $V_{MG}$ in (C-E) below. Color scale on $R_{xx}$ spans from 20~mK (blue) to 600~mK (red). (\textbf{C}) Average branch switching rate along the plateau. Error bars indicate uncertainty associated with the switching identification algorithm and background noise (see SM section 2). (\textbf{D}) Standard deviation of conductance computed for $G(t)$ at fixed $V_{MG}$. (\textbf{E}) The spectral densities of conductance plotted on a log-log scale (traces are arbitrarily vertically shifted by ordering of $V_{MG}$). Dotted lines are fit to the power law dependence $f^\alpha$ in the low frequency regime. Inset cartoons indicate the anyonic quasiparticle density of states versus energy, with the vertical dashed lines indicating the Fermi level for a given filling. (\textbf{F}) The exponent $\alpha$ extracted from the fits of the power spectrum within the frequency range associated with the RTN.}
    \label{fig3}
\end{figure*}
\begin{figure}[htbp]
    \includegraphics[width = 0.5\textwidth]{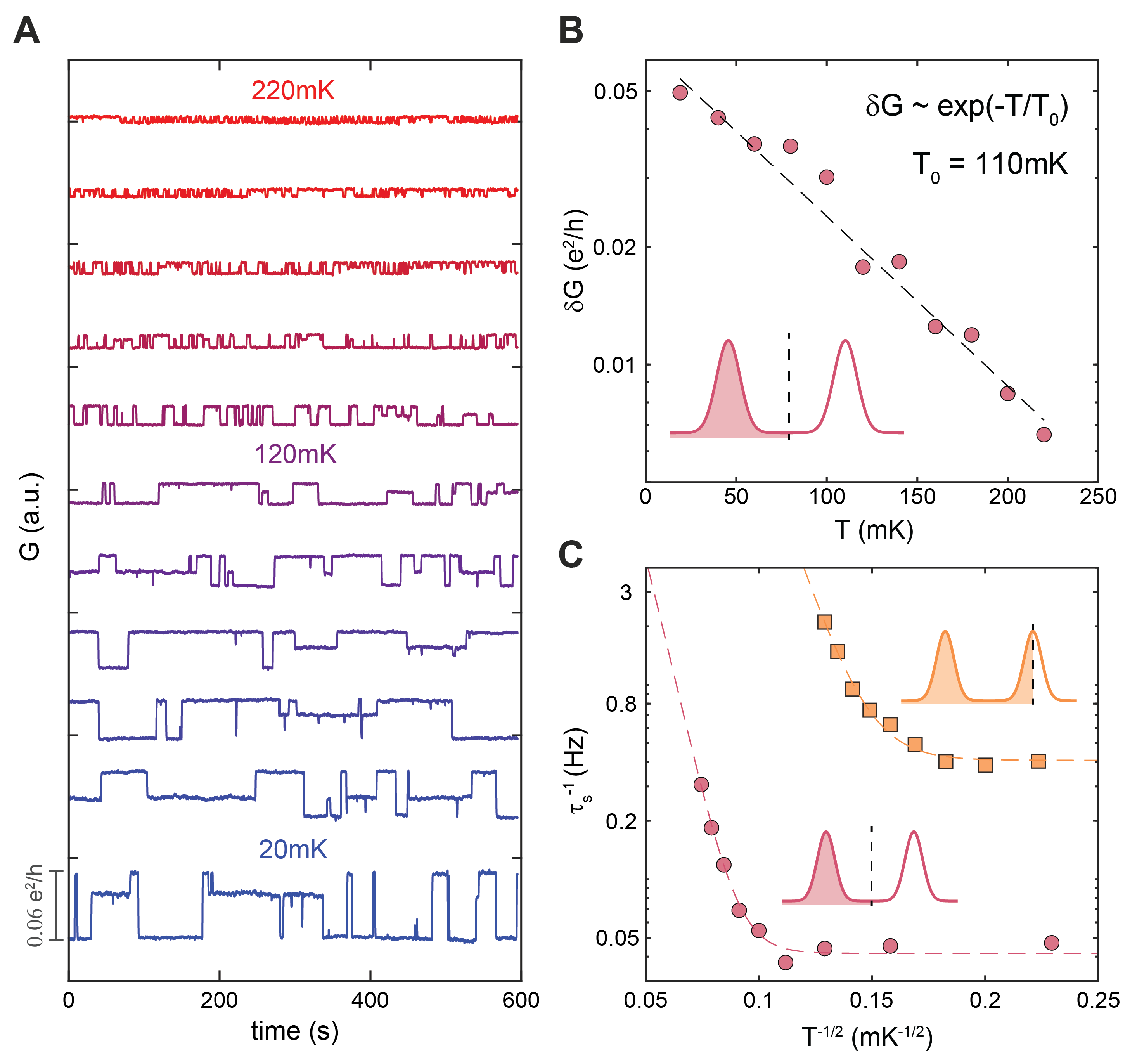}
    \caption{\textbf{Temperature dependent telegraph noise and visibility in $\bm{\nu = 1/3}$.} (\textbf{A}) Conductance versus time taken at different temperatures $T$ indicated by color (traces are arbitrarily vertically shifted by ordering of $T$). The 600~seconds shown are subsets of longer 50~minute datasets used in subsequent analysis. (\textbf{B}) Oscillation visibility extracted from $V_{PG}$ sweep measurements (21 repeated scans averaged per point). The exponential decay fit to the oscillation visibility results in a characteristic temperature $T_0 = 110$~mK. (\textbf{C}) Low-temperature saturation and high-temperature activation of the average switching rate $\tau_s^{-1}$ observed for two fillings: at the center and at the edge of the $\nu = 1/3$ plateau. Shown fit lines follow the empirical formula of Eqn. 2.}
    \label{fig4}
\end{figure}
Measuring how the magnetic field $B$ changes the total interference phase is crucial in determining if the interferometer falls into the AB or Coulomb-dominated regime \cite{halperin2011, feldman2021}. We start for comparison with the $\nu = 1$ integer QH state. By repeatedly sweeping the plunger gate, we observe the expected single-sinusoidal oscillations devoid of phase jumps (Fig. 2A). As we change $B$, we observe magnetic-field dependent AB interference (Fig. 2B), where the magnetic field period $\Delta B_1$ yields an area $\phi_0/\Delta B_1 = 0.80 \ \mu \text{m}^2$, as expected for interference of electrons in $\nu = 1$. A similar measurement in $\nu = 1/3$ with varying $B$ and  $V_{PG}$ produces an irregular 2D map  (Fig. 2C) since repeated plunger gate sweeps are not consistent due to the frequent RTN fluctuations. However, we observe that each plunger gate scan contains stochastic switching between 3 sinusoidal branches, again shifted by $2\pi/3$ (Fig. 2D). By superimposing 100 of these sweeps and plotting the histogram of total recorded conductance values over $V_{PG}$, we are able to sample enough instances of the AB oscillation signal of each branch to recreate the three interwoven sinusoids seen in the previous measurement technique discussed in Fig. 1. Fig. 2E exhibits the three interwoven branches at different magnetic fields. As the magnetic field decreases (from the top to bottom panels), we observe that each branch continuously drifts in the increasing $V_{PG}$ direction, creating a negative slope of constant phase (dashed line) analogous to the integer case of Fig. 2B. For $\nu=1/3$, the corresponding flux super-periodicity $\Delta B_{1/3}$ (the magnetic field period for a single branch to return to its initial $V_{PG}$ coordinate) yields $3\phi_0/\Delta B_{1/3} = 0.83 \ \mu \text{m}^2$. This value is what one expects for interfering $e/3$ quasiparticles such that $\phi_0 \rightarrow h/(e/3) = 3\phi_0$. We similarly observe AB interference and a flux super-periodicity on the fractional edge of $\nu = 4/3$ (Fig. S4), while the integer edge shows typical electron AB interference with no observable RTN.
\newpage
The fact that all tested configurations show AB interference (i.e. a negative slope of constant phase with respect to $V_{PG}$ and $B$) instead of Coulomb-dominated effects indicates that our graphene interferometer is sufficiently screened by the graphite gates to suppress long-range Coulomb interaction, as required for observing braiding signatures \cite{halperin2011, nakamura2020, nakamura2022}. Furthermore, the observed $3 \phi_0$ magnetic-flux super-period in both $\nu = 1/3$ and $\nu = 4/3$ agrees with the $3\phi_0$ super-period observed near the center of the $\nu = 1/3$ plateau in GaAs-based FP interferometers \cite{nakamura2020, nakamura2022}, suggesting that the quasiparticle gap is large enough to prevent continuous addition of quasiparticles \cite{rosenow2020}. 

\subsection*{Telegraph noise across the quantum Hall plateau}
We gain insight into the nature of the quasiparticle fluctuations by tracking how the RTN  evolves with density (filling) across the FQH plateau. Fig. 3A exhibits subsets of 50~minute measurements of the conductance, showing the $V_{MG}$ dependent RTN across the $\nu = 1/3$ plateau. The trace colors indicate the select values of $V_{MG}$ that correspond to the subrange of the $\nu = 1/3$ QH plateau as shown in Fig. 3B.
Of immediate note are clear changes in both the switching rate and amplitude as $V_{MG}$ steps across the plateau. The average switching rate from the complete 50 minute traces, $\tau_s^{-1}$, shown in Fig. 3C, exhibits steep rises near the edges of the plateau and a relatively constant switching rate in between. Fig. 3D shows the standard deviation, $\sigma_G$, of the diagonal conductance $G(t)$, providing a direct measure of switching amplitude.

The striking behavior of the anyon branch switching rate and amplitude along the plateau reveals that the quasiparticle fluctuations and coherence are closely correlated with the compressibility of the FQH state. A natural explanation would be to account for the increase in switching rate with an increase in the number of localized states, and thus anyon trapping sites, near the edges of the FQH plateau. However, a simple increase in the individual switching rates of a constant number of traps across the plateau cannot be ruled out with just this observation alone.

The noise spectral density of the diagonal conductance, $S_G$, can provide additional information supporting the hypothesis of an increasing trap number near the edge of the plateau. Fig. 3E shows $S_G$ as a function of frequency $f$, obtained from the Fourier transform of $G(t)$ at various $V_{MG}$ values across the plateau region. We find that $S_G (f)$ follows a power law scaling behavior ${\sim}f^\alpha$ in the low frequency regime corresponding to the RTN signal. Fig. 3F exhibits the scaling exponent $\alpha$ obtained from the line fits shown in Fig. 3E. In the middle of the plateau, the exponent $\alpha$ remains close to -2, which is expected from RTN with a single switching timescale \cite{weissman1988}. However, near the edges $\alpha$ sharply increases and deviates away from this stable value. It is known that when many instances of RTN with different switching timescales are convolved together, $S_G$ is expected to approach $1/f$, or $\alpha = -1$ \cite{kogan1996}. Accordingly, the change of $\alpha$ near the edges of the plateau suggests that there are more unique switching timescales at the edge of the plateau than in the middle. This observation, together with the switching rate behavior discussed above, supports that the increasing branch switching incidents across the FQH plateau are indeed associated with increasing localized anyon states in the bulk.

\subsection*{Temperature scaling}
We also study how the RTN changes as a function of temperature $T$. Fig. 4A shows $G(t)$ near the middle of the $\nu = 1/3$ plateau at $T$ ranging from 20~mK to 220~mK. As temperature increases, the switching amplitude decreases corresponding to a decrease in AB oscillation visibility. Fig. 4B shows separately measured AB oscillation visibility as a function of $T$ using plunger gate sweep measurements (similar to Fig. 2C-E). The visibility decreases following an exponential decay as $T$ increases, with a characteristic energy scale $T_0 = 110$~mK similar to that observed in $\nu = 1/3$ in GaAs \cite{nakamura2020}.

In addition to the decrease in visibility, the RTN also shows an increase in switching rate as temperature increases. Fig. 4C shows temperature dependent $\tau_s^{-1}$ for two representative data sets, one at the center (red) and the other at the edge (yellow) of the plateau. While switching rate indeed increases quickly at the high temperature limit, there is a dramatic flat saturation of $\tau_s^{-1}$ for both data sets at low temperatures, below about $100$~mK and $50$~mK for the plateau center and edge, respectively. We find that an empirical formula which combines a term motivated by an Efros-Shlovskii (ES) form of variable-range hopping (VRH) \cite{efros1975, ebert1983} with an additional temperature-independent transition rate, $\tau_{s,0}^{-1}$,
\begin{equation}
    \tau_s^{-1} = \tau_{s,0}^{-1} + \tau_{s,1}^{-1}(T) \ e^{-(T_{ES}/T)^{1/2}}
    \label{eqn2}
\end{equation}
can provide a reasonable fit to the data as shown by the dashed lines in Fig. 4C. Here $\tau_{s,1}^{-1}(T)\sim1/T$ and $T_{ES}$ is the characteristic ES VRH temperature scale. While VRH has been used in the past to model longitudinal conduction due to localized states in the quantum Hall effect \cite{ebert1983, bennaceur2012, giesbers2009, martin2004}, at the present moment the physical mechanism behind the temperature scaling of $\tau_s^{-1}$ here remains nebulous. We comment that there is a wealth of additional information present in the RTN signal, including how the probabilistic weightings of the phase branches (see the right inset of Fig. 1E) evolve with $T$ and $V_{MG}$, that could be analyzed in future studies to inform a microscopic model of the trapping sites present in FQH states.

While we were able to provide evidence that the anyon fluctuations are closely associated with the number of anyon localized states, and that they have consistent temperature scaling properties, it remains unclear whether they arise from device-specific considerations or rather from the fundamental nature of FQH states in monolayer graphene. Based on the uniformity of the AB-modulating sinusoids in Figs. 1F and 2E, we postulate that the anyons are neither hopping from nor getting trapped within regions close to the QPCs, as even small changes in environmental charge would drastically change their tunings and the resultant visibility of the interference signal. Additionally, although fluctuations have not been reported in the same manner in GaAs heterostructure-based fractional interferometry experiments \cite{nakamura2023, nakamura2022, nakamura2020} and more recently in bilayer graphene \cite{kim2024}, we note that the presence of slower or faster switching timescales could make initial observations and interpretation of anyon fluctuations challenging with standard DC measurement techniques.

\subsection*{Outlook}
In this work, we have shown that anyonic quasiparticle fluctuations and the resulting RTN in a FP interferometer can be used as a powerful tool to exhibit the complete set of braiding phases and exchange statistics present in abelian states. These techniques pave the way for future interferometer measurements of non-abelian anyon braiding in even-denominator FQH states, where the small energy gap may enhance the RTN and limit the range of AB phase modulation. Several theoretical studies have already discussed which signatures would be expected in such a case where RTN is present for non-abelian states \cite{kane2003, grosfeld2006, rosenow2012}. Moreover, our ability to observe the slow equilibration of anyons enables future experiments to probe nontrivial anyon dynamics and to dynamically control anyon number in similar devices. This is particularly relevant to the ultimate goal of constructing a topological qubit with non-abelian anyons, which requires an understanding of how to selectively control and modulate the quasiparticle number \cite{dassarma2005, stern2006, bonderson2006}.

\textit{Note added:} During the preparation of this manuscript, we became aware of a concurrent work using a similar graphene device~\cite{samuelson2024}.

\subsection*{Acknowledgments}
We thank Abhishek Banerjee, Yuval Ronen, and Steven H. Simon for helpful discussions. The major part of the experiment was supported by DOE (DE-SC0012260). J.R.E. acknowledges support from ARO MURI (N00014-21-1-2537) for sample preparation, measurement, characterization, and analysis. K.W. and T.T. acknowledge support from the JSPS KAKENHI (Grant Numbers 20H00354 and 23H02052) and World Premier International Research Center Initiative (WPI), MEXT, Japan. M.E.W. and A.Y. acknowledge support from Quantum Science Center (QSC), a National Quantum Information Science Research Center of the U.S. Department of Energy (DOE). B.I.H. acknowledges support from NSF grant DMR-1231319. Nanofabrication was performed at the Center for Nanoscale Systems at Harvard, supported in part by an NSF NNIN award ECS-00335765.

\textbf{Author contributions:} T.W., J.R.E., and P.K. conceived of and designed the experiment. T.W. and D.H.N. created the van der Waals heterostructure. T.W. performed the nanofabrication. T.W. and J.R.E. performed the measurements. M.E.W. and A.Y. provided the measurement cryostat and collaborated on discussions and analysis. K.W. and T.T. provided the hexagonal boron nitride crystals. T.W., J.R.E., and P.K. analyzed the data and wrote the manuscript with contributions from B.I.H.

\bibliography{main}

\end{document}

\title{Supplementary Materials: Anyon braiding and telegraph noise in a graphene interferometer}
\maketitle

% S1
\begin{figure}[p]
    \centering
    \includegraphics[width = 0.8\textwidth]{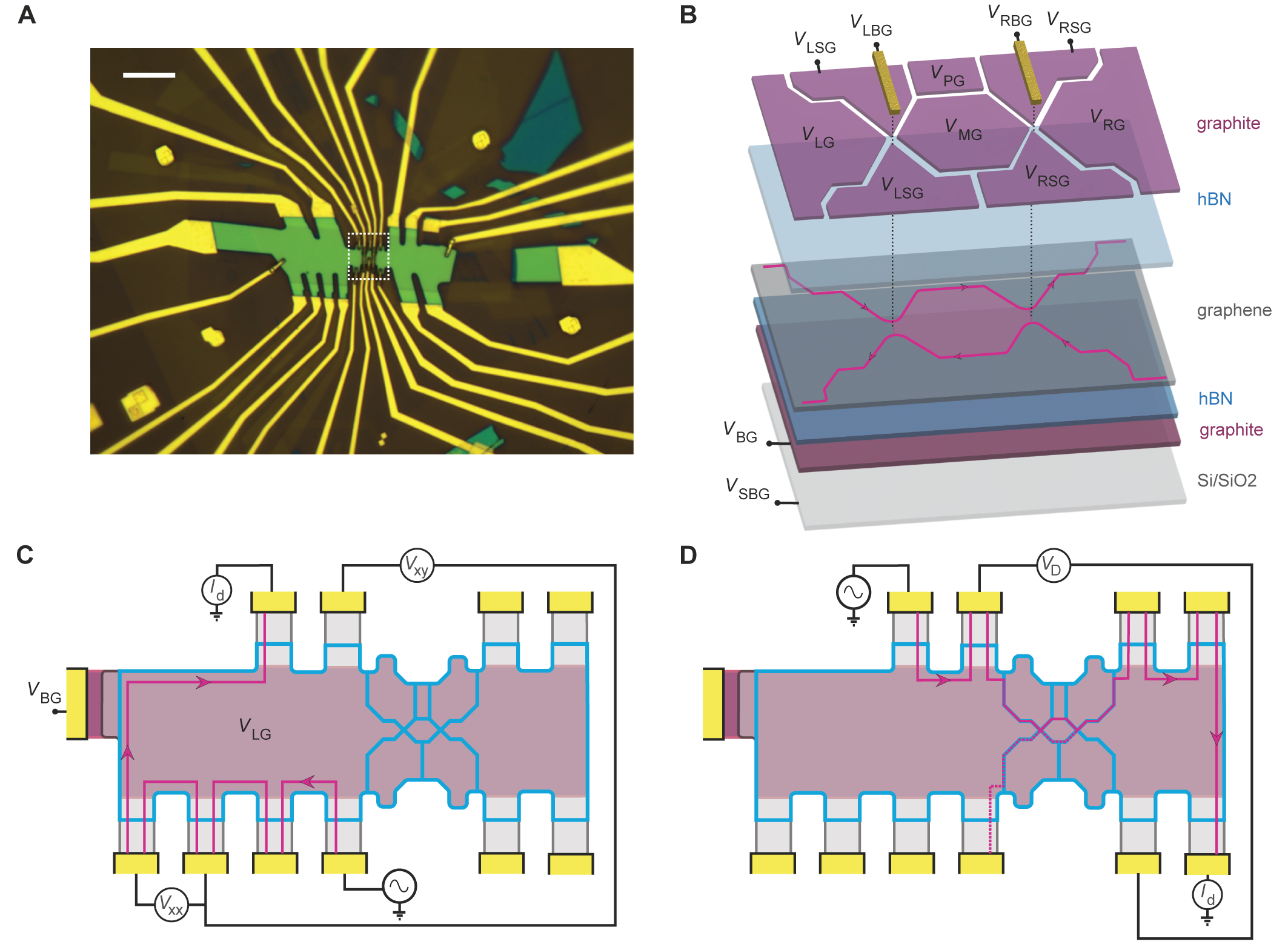}
    \caption{\textbf{Device contacts, gates, and measurement schemes.} (\textbf{A}) Optical micrograph of the device measured in this paper. Scale bar: 10 $\mu$m. (\textbf{B}) Schematic of the gating structure at the center of the interferometer i.e. zoomed in section outline by the dashed white line in (A). The bridge gates tune QPC transmissions by gating through the etched QPC gaps in the graphite top-gate. (\textbf{C}) Measurement schematic for quantum Hall transport measurements as reported in Fig. 1B-C and Fig. 3B. Here the signal source, drain, and chirality are such that the transport is confined to the left top-gate region. Hence, we tune density using only $V_{LG}$ and perform a standard Hall measurement to calibrate the expected filling factors for a given top-gate voltage. Blue lines indicate edges of the top-graphite gate and highlight the boundaries of the 8 separated regions. The purple region indicated the location of the graphite bottom-gate, while gray indicates the extent of the graphene layer. An ohmic contact to an extended graphite bottom-gate region allows an additional degree of freedom in tuning the QPCs by applying $V_{BG}$. Leads contact to the graphene channel in regions without a graphite top or bottom gate, necessitating the use of the silicon substrate to highly dope the region between the metal contact and channel. (\textbf{D}) Measurement schematic for QPC or interferometer conductance. Here the signal source and drain are on opposite sides of the QPCs, such that measurement of the ‘diagonal’ voltage drop $V_D$ yields the conductance of the QPC or interferometer as $G=I_d/V_D$  where $I_d$ is the measured drain current.}
    \label{s1}
\end{figure}
% S2
\begin{figure}[p]
    \centering
    \includegraphics[width = \textwidth]{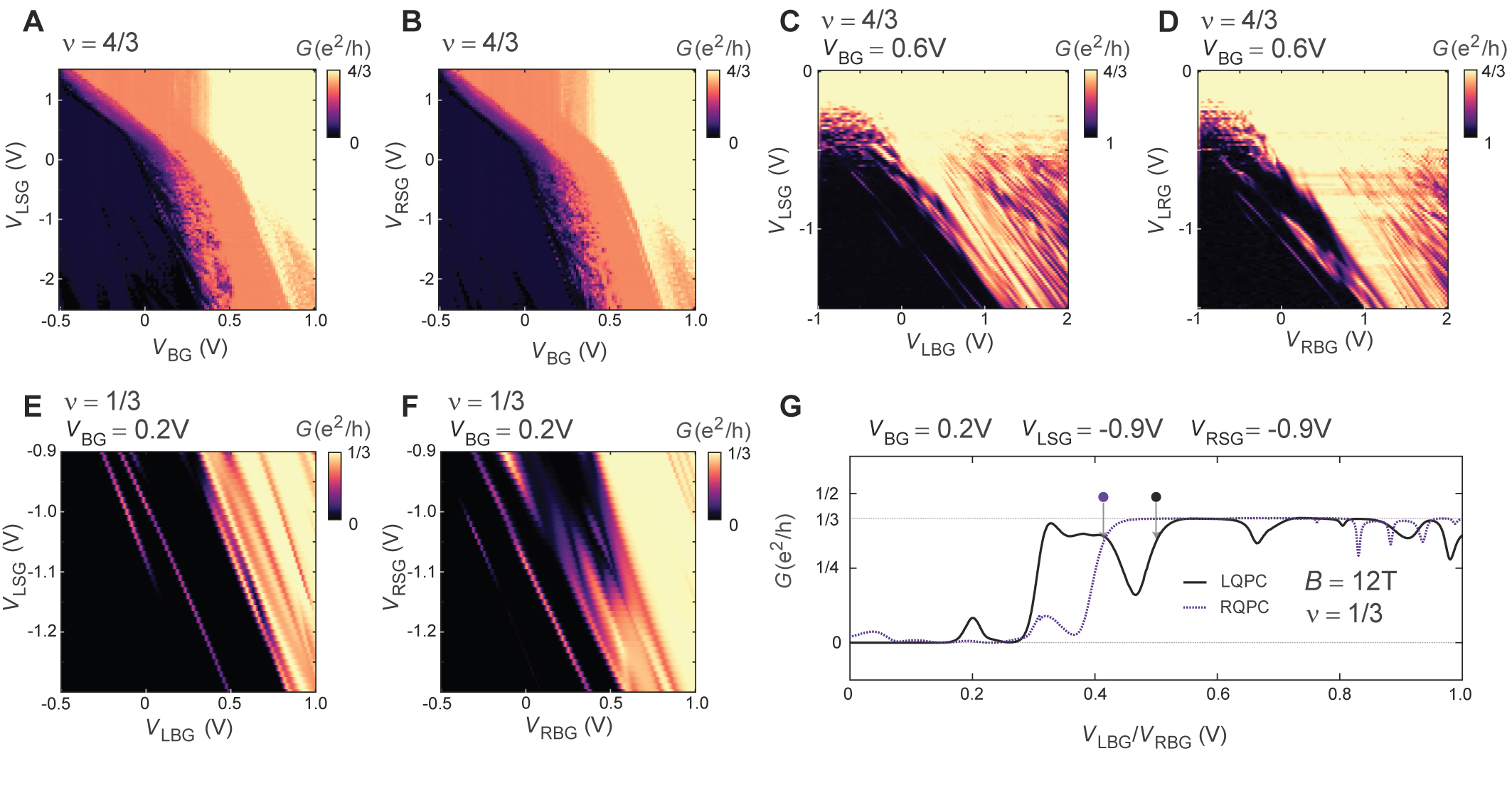}
    \caption{\textbf{QPC tuning details in $\bm{\nu = 4/3}$ and $\bm{\nu = 1/3}$.} (\textbf{A}) Conductance of the left QPC as a function of left split-gate voltage $V_{LSG}$ back-gate voltage $V_{BG}$ with the bulk of the device fixed at the center of the $\nu = 4/3$ plateau. (\textbf{B}) Conductance of the right QPC as a function of right split-gate voltage $V_{RSG}$ and back-gate voltage $V_{BG}$ at the same filling in $\nu = 4/3$. (\textbf{C}) Conductance of the left QPC with an appropriate back-gate voltage chosen as indicated to bias the QPC near an optimal operating point, which requires a larger density at the saddle-point to allow full transmission of both $\nu = 4/3$ edge channels in this case. With this back-gate voltage, we can now tune the transmission of the inner fractional edge channel using only the left split-gate and the left bridge-gate $V_{LBG}$, as shown, which simplifies further tuning. The outer integer channel is fully transmitted over this parameter space. (\textbf{D}) Same type of plot as (C) but for the right QPC. (\textbf{E}) Conductance of the left QPC with the bulk of the device fixed at the center of the $\nu = 1/3$ plateau and the indicated back-gate voltage. (\textbf{F}) Same type of plot as (E) for the right QPC. (\textbf{G}) Conductance through each QPC (left/right QPC tuned by $V_{LBG}$/$V_{RBG}$ respectively) with $\nu = 1/3$ set in the bulk and ($V_{BG}$, $V_{LSG}$, $V_{RSG}$) fixed as indicated. Black/purple dots indicate the left/right QPC tuning used to measure the data reported in the main text, corresponding to roughly 85\% transmission of the $\nu = 1/3$ edge channel.}
    \label{s2}
\end{figure}
% S3
\begin{figure}[p]
    \centering
    \includegraphics[width = 0.6\textwidth]{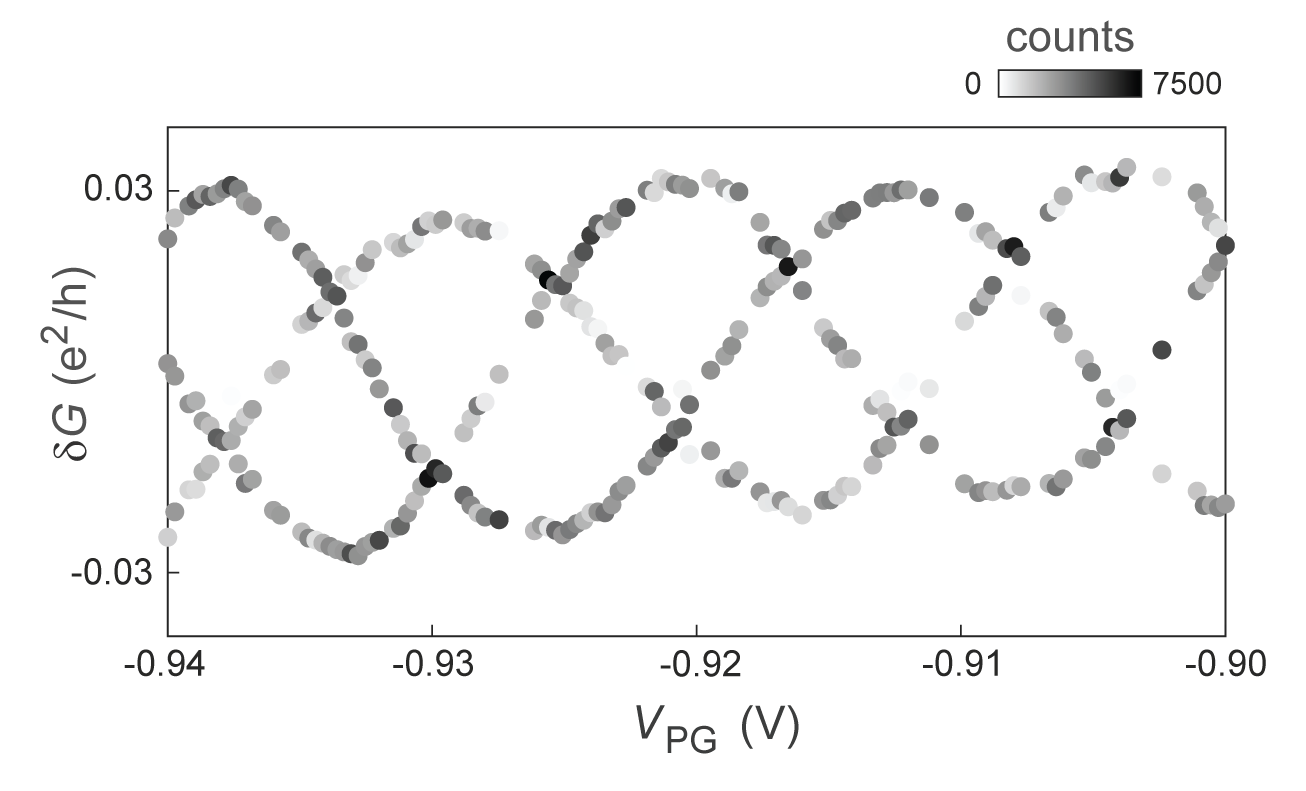}
    \caption{\textbf{Interwoven conductance branches extracted from static measurements in $\bm{\nu = 1/3}$.} Equivalent measurement in $\nu = 1/3$ to that shown in the main text for the fractional edge of $\nu = 4/3$  (Fig. 1E). Again, three phase-shifted branches are revealed weighted according to their total counts recorded at a sampling rate of 20 samples/s over an $\sim$8 minute time window for each $V_{PG}$ voltage.}
    \label{s3}
\end{figure}
% S4
\begin{figure}[p]
    \centering
    \includegraphics[width = 0.9\textwidth]{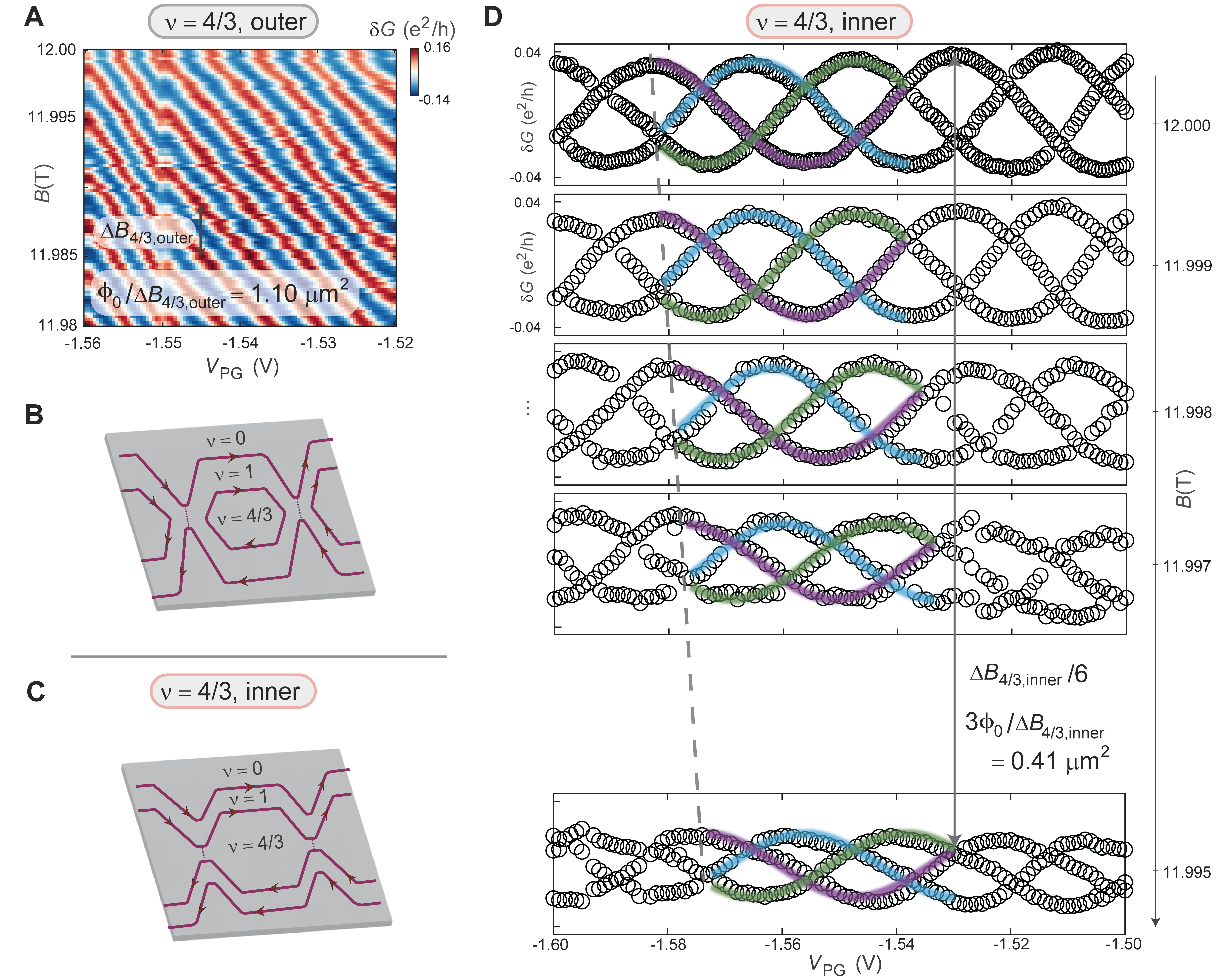}
    \caption{\textbf{Magnetic field dependence in $\bm{\nu = 4/3}$.} Equivalent measurement in $\nu = 4/3$ to that shown in main text for $\nu = 1/3$ (Fig. 2). Here, however, there is also an outer integer edge channel (i.e. $\nu = 1$) in addition to the inner fractional channel. Hence, we can tune to interfere either the outer integer or the inner fractional channel while holding the bulk filling $\nu = 4/3$ fixed. (\textbf{A}) Magnetic field dependence of conductance oscillations with $V_{PG}$ while the outer integer edge channel is partitioned at both QPCs. We observe a typical AB interference pattern with a magnetic field period yielding an enclosed area 1.1 $\mu \text{m}^2$ which matches the lithographic design. Notably, we do not observe telegraph noise, which enforces that interference of fractional quasiparticles is required to observe the phase shift due to anyon braiding. (\textbf{B}) Schematic showing interference of the outer integer edge channel, which flows at the boundary of the $\nu = 0$ and $\nu = 1$ incompressible regions. (\textbf{C}) Schematic showing interference of the inner fractional edge channel, which flows at the boundary of the $\nu = 1$ and $\nu = 4/3$ incompressible regions. (\textbf{D}) Magnetic field dependent panels extracted using the method in Fig. 1E/F, where we plot the three mean conductance values for each $V_{PG}$ without a weighting, for clarity. The interwoven sinusoidal oscillations trend in the same AB direction with decreasing magnetic field (downward in this plot) and exhibit a periodicity consistent with a $3\phi_0$ flux super-period. The flux period is consistent with the fractional edge enclosing an area of 0.41 $\mu \text{m}^2$, which is plausible since the area enclosed by the inner channel should be smaller than the area enclosed by the outer channel due to the width of the incompressible/compressible regions.}
    \label{s4}
\end{figure}
% S5
\begin{figure}[p]
    \centering
    \includegraphics[width = 0.9\textwidth]{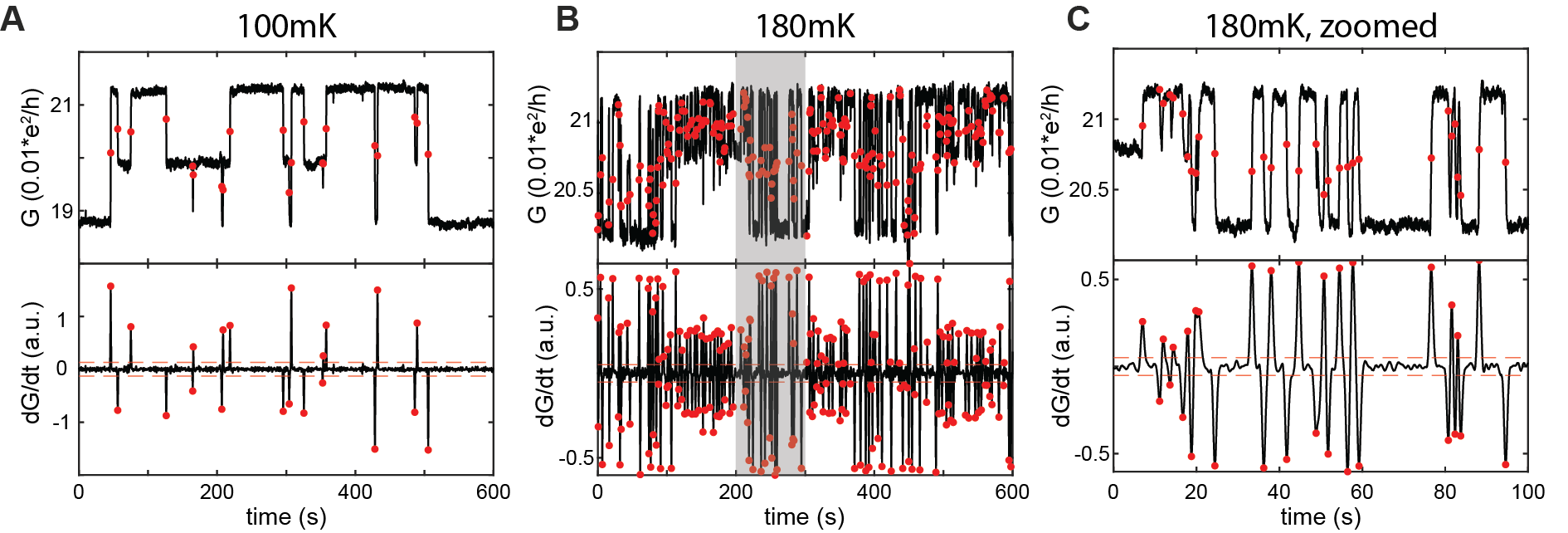}
    \caption{\textbf{Branch switching detection algorithm.} (\textbf{A}) Raw conductance data and derivative of the RTN signal taken within the middle of the $\nu = 1/3$ plateau at $T = 100$~mK, $B = 12$~T. The output of the switch detection algorithm described in the main supplement text is indicated by the red dots. The red dashed line on the derivative signal shows the peak finding threshold value ($\sqrt{2}$ times the derivative’s peak-to-peak noise floor). (\textbf{B}) RTN signal taken at the same device configuration but at 180~mK. Grey box indicates zoom window shown in \textbf{(C)}.}
    \label{s5}
\end{figure}
% S6
\begin{figure}[p]
    \centering
    \includegraphics[width = 0.4\textwidth]{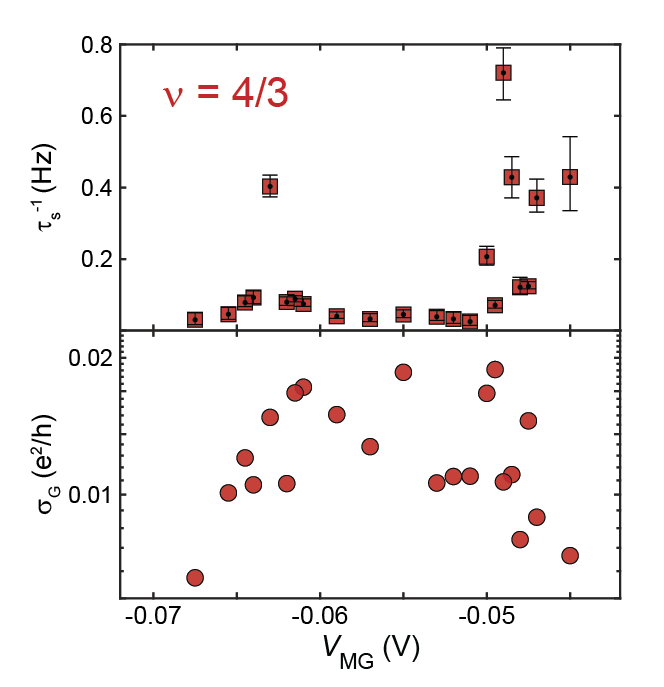}
    \caption{\textbf{Plateau analysis of RTN for $\bm{\nu = 4/3}$.} \textbf{Top:} Average branch switching rate along the $\nu = 4/3$ plateau, analogous to Fig. 3C of the main text which shows $\nu = 1/3$. \textbf{Bottom:} Standard deviation $\sigma_G$ of conductance, analogous to Fig. 3D of the main text.}
    \label{s6}
\end{figure}
% S7
\begin{figure}[p]
    \centering
    \includegraphics[width = 0.9\textwidth]{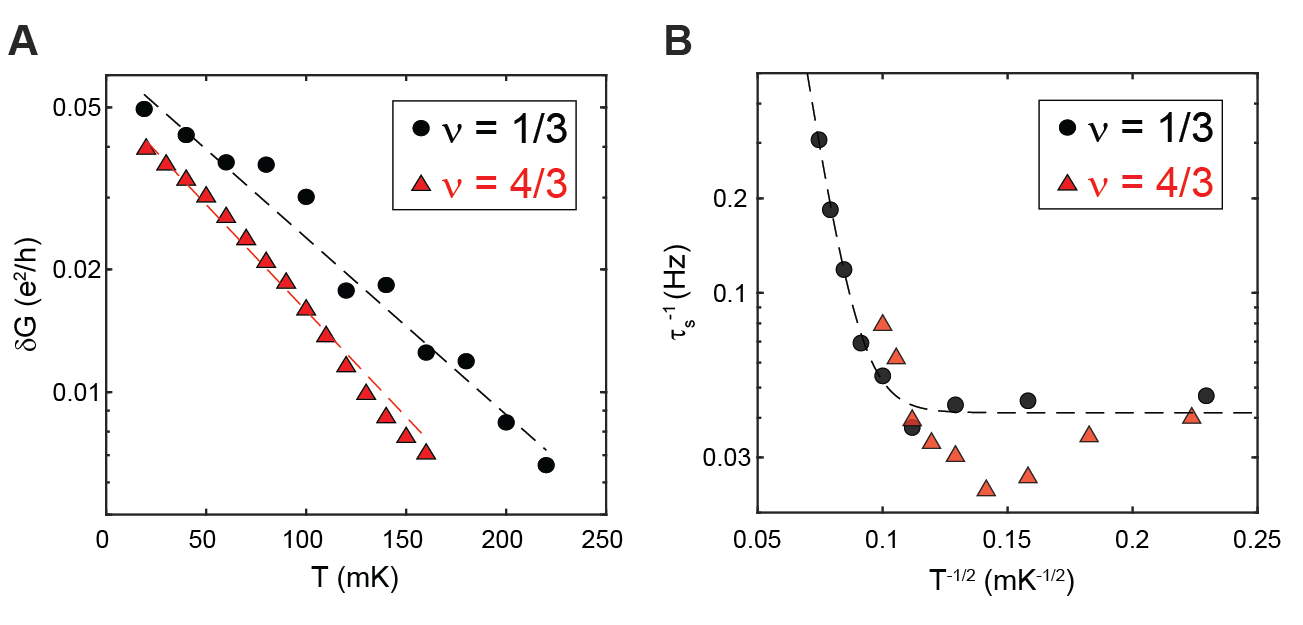}
    \caption{\textbf{$\bm{\nu = 4/3}$ RTN temperature scaling.} (\textbf{A}) Oscillation visibility extracted from $V_{PG}$ sweep measurements for $\nu = 1/3$ (same data shown in main text Fig. 4B) and $\nu = 4/3$, both taken with $V_{MG}$ set to be in the middle of each filling factor’s respective plateau. (\textbf{B}) Average switching rate for each filling factor ($\nu = 1/3$ data copied from Fig. 4 of main text). The $\nu = 4/3$ data is only taken up to 100 mK due to the noise floor overwhelming the separation between two of the three conductance plateaus.}
    \label{s7}
\end{figure}

\section*{Materials and Methods}
\subsection{Device Fabrication}
Creating a graphene-based device with low enough charge disorder to realize robust fractional quantum Hall states requires special considerations in the fabrication process. This affects primarily the order of the fabrication steps, since the large number of electron beam lithography (EBL) and reactive ion etching (RIE) steps tend to introduce disorder via surface residues of the polymethyl methacrylate (PMMA) resists that we use as masks. Therefore, to screen out these residue by-products of the fabrication steps from the graphene two-dimensional electron system (2DES), we begin by encapsulating the graphene channel in top and bottom insulating hexagonal boron nitride (hBN) and conducting graphite (generally $\sim$5 nm thick) layers. The hBN layers serve as a near-atomically flat dielectric through which the graphite layers serve as metallic gates. The device reported here consists of a top/bottom hBN thickness of 49/27 nm and has previously been measured extensively in integer quantum Hall states [28]. 

These thin dielectric layers serve a dual purpose of bringing the conducting graphite layers close to the graphene, which tends to screen interactions within the graphene 2DES. We believe that this is crucial for observing Aharonov-Bohm oscillations in the device, as similar graphene-based devices without close screening layers display Coulomb-dominated oscillations [21]. Likewise, GaAs devices approaching the small areas of our device displayed Coulomb-dominated oscillations [44, 45]. It required special sample growth utilizing screening wells to screen out interactions sufficiently strongly to drive the device into the Aharonov-Bohm regime and observe fractional interference, which was only achieved in the most recent devices [14-16].

We fabricate the 5 layer van der Waals stacks (Fig. S1) using the now standard polycarbonate (PC) dry transfer method described in detail in our previous work [20]. When the 5 layers are picked up on the PC stamp, we press the stack onto a doped Si substrate with 300 nm SiO2 and remove bubbles in the stack using repeated re-lamination at elevated temperatures and tilt angle. Melting the PC at 180C now allows us to release the clean stack onto the substrate. Subsequently, we dissolve PC residues in chloroform for 12 hours and perform a 3 hour vacuum anneal at 300C to strongly adhere the stack to the substrate. We then perform the same steps of EBL, RIE, and deposition to construct the device as detailed previously [20]. We additionally add final bridge gates over the two QPCs in this device to separately tune transmissions. The final device is shown in Fig. 1 and Fig. S1

\subsection{Measurement Details}
In addition to the need for a low-disorder device, an interference experiment in the fractional quantum Hall regime requires extremely low electron temperatures for large enough phase coherence of the interfering quasiparticle wavefunction (note the temperature scale $T_0=110$~mK for exponential decay of oscillation visibility in $\nu = 1/3$ in Fig. 4B), which necessitates measurements in a well-tuned dilution fridge and, crucially, carefully designed electronic filtering. We performed the experiments in an Oxford MX400 dilution fridge with base temperature of 18~mK. All DC transport lines were thermalized through Thermocoax cables and 3 Sapphire plates between room temperature and the mixing chamber before breaking out into copper wire to feed into a custom passive component filter PCB assembly. This circuit consists of three Mini-Circuits LFCN-xxxx+ low pass filters (stepped down in frequency cutoff from 5000 MHz-1800 MHz-80 MHz) followed by a single-pole RC filter (R = 10 k$\Omega$, C = 3.3 nF). While the engineered RC cutoff is ~4.8 kHz, the practical cutoff is significantly lower (we estimate a cut-off as low as 50 Hz when the sample resistance is high) due to the device’s many contacts to the same 2DEG, allowing the capacitor of the measurement lines to be weighted by the others seen through the device’s resistance. The PCB uses 370HR as the dielectric base material and the trace finish coating uses an immersion silver process with no strike layer. An exposed ring of the ground plane is clamped to a soft gold-coated copper enclosure that is directly connected to the mixing chamber copper chassis. The enclosure also acts as a Faraday cage around the PCB components. This thermalizing and filtering scheme ensures low electron temperature likely within a few mK of the fridge base temperature, as indicated by interferometer visibility scaling (see Fig. 4B).

 Most top graphite gate regions and the bridge gates are controlled with a home-made 16-bit DACs, while the silicon backgate is supplied 40 V using a Keithley 2400. Special care is taken to ensure that the middle gate ($V_{MG}$), which sits above the interferometer cavity and sets its density, as well as the plunger gate ($V_{PG}$), are as electrostatically stable as possible to ensure high interferometer visibility and minimal dynamical effects. To this end, these two gates are controlled with additional custom built 20-bit DACs based around the AD5791 integrated circuit. Each channel uses its own ovenized Zener diode voltage reference circuit that has been run continuously for over 3 months with efforts made to thermally insulate it from the rest of the circuitry and environment. Based on our measurements, this results in an overall DAC output drift of $<0.6$ LSB/day (equivalent to $<11 \mu $V/day). The output noise floor of the DAC is $<25$ nV/$\sqrt{\text{Hz}}$ within a 100 kHz measurement bandwidth.

We performed the measurements using typical low-frequency lock-in amplifier techniques using SR830 lock-in amplifiers, SR560 voltage preamps, and Ithaco 1211 current preamps. For interferometer and QPC measurements, we typically voltage bias using the SR830 connected to a voltage-divider to apply $15 \ \mu$V to the sample, resulting in current of order 100 pA to minimize reduction of the oscillation visibility due to finite-bias effect. We measure Hall data using 1 mV bias at 1.0777 Hz to minimize capacitive contribution and increase the resulting measurement accuracy of $V_{xx}$ and $V_{xy}$. In contrast, the majority of the interference data was measured at 73.77 Hz, allowing fast enough sampling ($\sim$20 samples/s) to resolve the telegraph noise events. However, due to filter attenuation, the amplitude of the signal was partially attenuated at this frequency. As we only consider conductance oscillations due to the interference, i.e. $\delta G$ on top of the mean conductance, this attenuation does not affect our results and conclusions. Lastly, to achieve accurate QPC calibrations (Fig. S2), where we need accurate values of $G$, as well as when we measure oscillation visibilities, we instead use an intermediate frequency of 11.117 Hz that reduces measurement time while getting accurate enough values.

\section*{Supplementary Text}
\subsection*{Section 1 - Device tuning details}
We show an optical image of our device and schematics of the gating and measurement schemes in Fig. S1. The 8 separated top graphite regions (Fig. S1B) are separately tuned to electrostatically define each QPC as described later. We additionally employ a global bottom graphite gate, which can set an overall density offset at the QPC saddle points, where the top graphite has been etched away at a width of $\sim$150 nm. The density offset is necessary to allow full transmission of edge channels at the QPCs while the side split-gates deplete electrons to maintain $\nu = 0$. Finally, we typically apply +40 V to the doped Si substrate to dope the graphene channel contacts through the 300 nm surface oxide and bottom hBN layers. This contact doping is necessary since we contact the channel at regions that extend outward from both the top and bottom graphite gates. In Fig. 1C we show the arrangement of the gates from a top-down view and which contacts are used for a typical Hall measurement. We perform these calibration measurements within a single region of the device (tuned by $V_{LG}$) to get the most accurate measurements. Lastly, we show the typical configuration for all measurements of QPC and interferometer conductance reported in this work in Fig. S1D.

Our process to tune the QPC transmissions for both $\nu = 4/3$ and $\nu = 1/3$ is shown in Fig. S2. We begin (Fig. S2A) by taking a large range parameter sweep of the left QPC split gate $V_{LSG}$ vs. the global bottom-gate $V_{BG}$ with the filling factor in the other regions of the device fixed to $\nu = 4/3$  (this required sweeping $V_{RSG}$, $V_{LG}$, $V_{MG}$, $V_{PG}$, and $V_{RG}$ simultaneously to maintain constant density as $V_{BG}$ sweeps). We show a parameter sweep for the right QPC in Fig. S2B. These scans enable us to distinguish the regions in which the split-gates are in $\nu = 0$ or below, hence reducing conductance to zero along a line controlled predominantly by $V_{LSG}$/$V_{RSG}$, from regions in which QPC saddle point transitions from zero to full conductance, controlled predominantly by $V_{BG}$. Informed by these scans, we then set $V_{BG}$ close to the saddle point transition from $\nu = 1$ to $\nu = 4/3$ (and set $V_{RSG}$, $V_{LG}$, $V_{MG}$, $V_{PG}$, and $V_{RG}$ accordingly to $\nu = 4/3$) and perform another parameter sweep now of $V_{LSG}$ and $V_{LBG}$ in Fig. S2C. We show similar data for the right QPC in Fig. S2D. QPC tunings shown in Fig. 1D were chosen from this data set such that the fractional edge mode could be directly tuned from full reflection to full transmission using only $V_{LBG}$/$V_{RBG}$ while holding all other gates fixed. We similarly show QPC parameter scans for  $\nu = 1/3$ in Fig. S2E/F and the resulting tunings chosen for the interference measurements performed in $\nu = 1/3$ in Fig. S2G.

The particular voltages ($V_{BG}$, $V_{LSG}$, $V_{RSG}$) are chosen so that a simple scan of $V_{LBG}$/$V_{RBG}$ can tune the left/right QPC from full reflection to full transmission with a minimum of transmission resonances (i.e. sharp, reproducible conductance oscillations) in between. Resonances are believed to be caused by random quantum dot sites near the QPC saddle point, which can be pinned by disorder and which strongly affect the conductance as they tune through a resonant condition [43]. By tuning to a sharper confining potential, the effect of these resonances and edge state reconstruction can be reduced [26]. Hence, we generally attempt to increase the sharpness of the confining potential in our tunings by choosing gate voltages that increase the electric field near the boundary, practically meaning we tune to more negative split-gate voltages if possible.

\subsection*{Section 2 - Algorithm used for determining RTN switching events}
To determine when a RTN switching even between conductance plateaus occurs, we use the following algorithm: The derivative of the raw diagonal conductance data is taken and then smoothed with a gaussian-weighted average over a window of 30 data points. The smoothing is performed in order to reduce the background noise to allow for fitting of faster switching events to ensure accurate and consistent sampling across scans. Then, all peaks above a threshold magnitude are found and recorded. The threshold value is set to be $\sqrt{2}$ times the peak-to-peak noise floor of the raw conductance signal’s derivative, which is an arbitrary yet consistent decision across all data. The background noise appears to be sensitive to temperature and not the value of $V_{MG}$ (i.e. location within the filling factor plateau). Consequently, the threshold value is identical for the data presented in Fig. 3C and is scaled for that shown in Fig. 4C. Note that the final output of the algorithm is cross-checked with the raw conductance data as shown in Fig. S5 ensuring accuracy and consistency in picking out switching events and disregarding noise spikes.